\documentclass[11pt]{article}
\usepackage{amsmath, amssymb, amsthm}
\usepackage{graphicx}
\usepackage{amsmath}
\usepackage{epsfig}
\usepackage{url}
\usepackage{algorithmic}
\usepackage{fullpage}

\newtheorem{definition}{Definition}[section]

\title{An Experimental Comparison of PMSprune and Other Algorithms for Motif Search
\thanks{This work has been supported in
part by the following grants: NSF 0326155, NSF 0829916 and NIH
1R01GM079689-01A1.}}

\author{Dolly Sharma \\
University of Connecticut\\
\textsf{dolly@engr.uconn.edu} \and
Sanguthevar Rajasekaran\\
University of Connecticut\\
\textsf{rajasek@engr.uconn.edu} \and
Hieu Dinh\\
University of Connecticut\\
\textsf{hdinh@engr.uconn.edu} \and}

\begin{document}

%\pagenumbering{roman}  % Roman numerals from abstract to text
\maketitle              % you need to define \title{..}
\thispagestyle{empty}

\begin{abstract}

Extracting meaningful patterns from voluminous amount of biological
data is a very big challenge. Motifs are biological patterns of
great interest to biologists. Many different versions of the motif
finding problem have been identified by researchers. Examples
include the Planted $(l, d)$ Motif version, those based on
position-specific score matrices, etc. A comparative study of the
various motif search algorithms is very important for several
reasons. For example, we could identify the strengths and weaknesses
of each. As a result, we might be able to devise hybrids that will
perform better than the individual components. In this paper we
(either directly or indirectly) compare the performance of PMSprune
(an algorithm based on the $(l, d)$ motif model) and several other
algorithms in terms of seven measures and using well established
benchmarks

In this paper, we (directly or indirectly) compare the quality of
motifs predicted by PMSprune and 14 other algorithms. We have
employed several benchmark datasets including the one used by Tompa,
et.al. These comparisons show that the performance of PMSprune is
competitive when compared to the other 14 algorithms tested.

We have compared (directly or indirectly) the performance of
PMSprune and 14 other algorithms using the Benchmark dataset
provided by Tompa, et.al. It is observed that both PMSprune and DME
(an algorithm based on position-specific score matrices) in general
perform better than the 13 algorithms reported in Tompa et. al..
Subsequently we have compared PMSprune and DME on other benchmark
data sets including ChIP-Chip, ChIP-seq, and ABS. Between PMSprune
and DME, PMSprune performs better than DME on six measures. DME
performs better than PMSprune on one measure (namely, specificity).

\end{abstract}

\newpage
\pagenumbering{arabic}

\section{Introduction}\label{sec:intro}
Genomic sequencing allows the identification of nearly all of the
genes in an organism, from which the potential gene products can be
inferred. However, genomic sequencing does not provide direct
information on the function of the individual gene products, nor
their interrelationships.  Motifs provide clues on this function.
Several computational approaches for identifying motifs can be found
in the literature. These techniques differ in how motifs are defined
and modeled. Each approach looks at a different facet of motifs. No
single model or technique can identify all possible motifs. The
final verification, as to whether a motif is functional or not, can
only be assessed experimentally. Since experiments are expensive and
time consuming, computational techniques that can select potential
functional motifs with high accuracy are desirable.

Many versions of motif search have been identified in the
literature. Examples include planted $(\ell, d)$ motif search (PMS),
edit distance based motif search (EMS), and simple motif search
(SMS). These algorithms have a common theme. In particular, they
look for substrings that are common to many of the given input
sequences. Algorithms based on these models have proven to be quite
useful in practice for solving problems such as the identification
of transcription factor binding sites. PMSprune \cite{ref03} is an
algorithm based on the $(\ell, d)$-motif model that is currently
known to solve the largest challenging instances. Another important
class of motif finding algorithms is based on position-specific
score matrices. These algorithms are based on identifying the
statistical difference between a substring and the background.
Numerous papers have been written for motif search based on these
two and other techniques. It will be highly desirable to compare the
various motif search algorithms to identify the advantages and
drawbacks of each technique. This comparative study could lead to
the discovery of better algorithms. A study of this kind has been
conducted by Tompa, \emph{et. al.} \cite{ref12} who compared the
performance of 13 different algorithms based on seven measures. In
this paper we add the comparison of two more algorithms to this
list. These two algorithms are PMSprune \cite{ref03} and DME
\cite{ref13} (an algorithm based on the position-specific score
matrices).

\textbf{The $(l, d)$-motif Problem (LDMP):} This problem takes as
input $n$ sequences each of length $m$. Input also are integers
$\ell$ and $d$. The problem is to find a motif $M$ (a string of
length $\ell$) that occurs in each input sequence up to hamming
distance of $d$.

Numerous algorithms have been developed to solve LDMP. LDMP
algorithms can be categorized into two, namely, exact algorithms and
approximation algorithms. An exact algorithm (also called an
exhaustive enumeration algorithm) always finds the correct
answer(s). On the other hand, an approximation algorithm (also known
as a heuristic algorithm) may not always output the correct
answer(s). Some examples of algorithms proposed for LDMP are Random
Projection \cite{ref05}, MITRA \cite{ref06}, Winnower \cite{ref07},
Pattern Branching \cite{ref08}, PMS1, PMS2, PMS3 \cite{ref01}, PMSP
\cite{ref09}, CENSUS \cite{ref21} and Voting \cite{ref11}. Out of
these algorithms, CENSUS, MITRA, PMS1, PMS2, PMS3, PMSP and Voting
are exact algorithms.

Certain instances of LDMP have been identified to be challenging
instances. In particular, an $(\ell, d)$-motif instance is said to
be challenging if the expected number of motifs that occur in the
input by random chance is greater than or equal to one (when the
input sequences themselves are generated randomly). For example, the
instances (9,2), (11,3), (13,4), and (15,5) are examples of
challenging instances. It is customary in the literature to show the
performance of LDMP algorithms only on challenging instances. It is
always a challenge to solve as large a challenging instance of LDMP
as possible.

Pevzner and Sze \cite{ref07} have proposed the WINNOWER algorithm.
It generates a list of all the $l$-mers present in the database and
constructs a graph out of them. Each $l$-mer represents a node and
there is an edge between two nodes if the hamming distance between
these two nodes is at most $2d$ and they belong to different
sequences. The problem of finding motifs is then reduced to finding
large cliques in this graph. The problem of finding a maximum clique
in a graph is known to be NP-hard. The authors use some novel
pruning techniques to find large cliques. WINNOWER uses a technique
to generate Extendable Cliques. It is an iterative algorithm. The
run time of this algorithm is $O(N^{2d+1})$, where $N=nm$.  SP-STAR
uses a technique which eliminates more edges than WINNOWER and hence
is faster. It uses less memory as well.

PatternBranching \cite{ref08} performs a local search. It generates
the neighbors of all $\ell$-mers present in the sequences and uses a
scoring method to assign scores for these neighbors. There are
$n(m-l+1)$ $\ell$-mers presenting in the sequences. Each $\ell$-mer
has ${\ell \choose d}3^d$ neighbors. The scores of all the neighbors
of all the $\ell$-mers are computed and the best scoring neighbors
are identified. Since it searches through only a selected set of
$\ell$-mers, it is very efficient. This algorithm takes a $u$, and
then computes the best neighbor $u_1$ of $u$. In the next step it
identifies the best neighbor $u_2$ of $u_1$ and so on till it
computes $u_d$. The $u_d$ values for all possible $u$ are computed
and then the best $u_d$ value becomes the output. In every step it
keeps only one value of the best neighbor but there could be a set
of $\ell$-mers which could be the best neighbors for $u$.

MITRA \cite{ref06} is based on WINNOWER \cite{ref07}. It uses pair
wise similarity information. It uses a mismatch tree data structure
and splits the space of patterns into subspaces which start with a
given prefix. Pruning is applied to each of these subspaces. MITRA
performs very well in practice and is one of the best performing
algorithms designed for Planted Motif Search Problem.

AlignACE \cite{ref14} uses the whole genome RNA quanititation to
identify genes that are most responsive to environmental or
genotypic change. Searching for mutually similar DNA elements among
the upstream non-coding DNA sequences of these genes, the candidate
regulatory motifs and corresponding candidate sets of co-regulated
genes can be identified. This technique was applied to three
regulatory systems in yeast accharomyces cerevisiae: galactose
response, heat shock, and mating type. ANN-Spec \cite{ref15} uses an
Artificial Neural Network and a Gibbs Sampling Method to define the
Specificity of a motif. It searches for parameters that will
maximize the Specificity of the Binding sequence compared to the
Background.

GLAM \cite{ref16} adds some useful enhancements to the Gibbs
Sampling alignment method like automatic detection of alignment
width, calculation of statistical significance and optimization of
alignment by Simulated Annealing. The Improbizer \cite{ref17}
searches for motifs in DNA or RNA sequences that occur with
improbable frequency (to be just chance) using a variation of the
expectation maximization (EM) algorithm. Oligo/dyad Analysis
\cite{ref19} detects statistically significant motifs by counting
the number of occurrences of each word or dyad and comparing these
with expectation. The most crucial parameter is choosing the
probabilistic model for the estimation of occurrence significance.
In this study, a negative binomial distribution on word
distributions was obtained from 1000 random promoter selections of
the same size as the test sets.

QuickScore \cite{ref20} uses an extended consensus method allowing
well defined mismatches and uses mathematical expressions for
computing $z$-scores and P values. SeSiMCMC \cite{ref21} consists of
two stages. In the first stage the weight matrix for a given motif
and spacer length is optimized and in the second stage looks for the
best motif and spacer lengths for obtained motif positions. It
optimizes the common information of the motif and of distributions
of motif occurrence positions.

Weeder \cite{ref22} is a consensus-based method where each motif is
evaluated according to the number of sequences it appears in and
also on how well-conserved they are in these sequences. YMF
\cite{ref23} performs an exhaustive search to find motifs with
greatest $z$-scores. A $p$-value is used to assess significance of
the motif.

Several algorithms for PMS have been proposed in \cite{ref01}. These
are exact algorithms based on radix sorting. The first algorithm
proposed was PMS1. PMS1 simply generates neighbors of all the
$\ell$-mers present in a sequence and saves it in an array $C_i$,
where $i$ is the sequence number. It repeats this for all the
sequences. The arrays are then sorted. The $\ell$-mers we get with
the intersection of all $C_i$'s, where $1 \leq i \leq n$, is the
list of motifs. However, PMS1 is not capable of solving challenging
instances with large values of $\ell$. Some improvements have also
been proposed on PMS1 in \cite{ref01}. Another algorithm proposed is
PMS2 \cite{ref01}, which is capable of solving challenging instances
up to $d = 4$. For instance, it could not solve the Challenging
instance (15, 5). The third algorithm proposed in \cite{ref01} was
PMS3 which was later implemented and is capable of solving the
challenging Instance (21, 8). Two new algorithms PMSP and PMSi were
proposed in \cite{ref03} which were built upon PMS1 and perform a
lot better than PMS1.

PMSP \cite{ref03} uses the idea of exploring the neighbors  of all
the $\ell$-mers that occur in the first sequence and then checking
if these represent a motif or not. The worst case time bound of this
algorithm is worse than that of PMS1 but in practice it performs
much better. It is capable of solving the challenging instance (17,
6).

PMSPprune \cite{ref03} is another algorithm based on PMS1 and is an
improvement over PMSP. It uses a branch and bound approach. It shows
significant improvements in terms of time taken and is also capable
of solving the challenging instance (19, 7). In this paper we use
qPMSprune, which is a special case of PMSprune to perform our tests.

DME (Discriminating Matrix Enumerator) was proposed in \cite{ref13}.
It uses an enumerative algorithm to exhaustively and efficiently
search a discrete space of matrices. Each matrix is scored according
to its relative over representation. A local search procedure is
used to refine the highest-scoring matrices, which optimizes the
relative over representation score. As soon as a motif is
discovered, its occurrences are erased from the data and the
procedure is repeated to discover additional motifs. In Tompa, et
al. \cite{ref12} 13 different motif finding algorithms have been
tested with respect to 7 different statistical measures. We have
tested the performance of PMSprune and DME over the benchmark
datasets provided by Tompa, \emph{et.al.} \cite{ref12}. It is
observed that the performances of PMS and DME are in general better
than the 13 algorithms compared in \cite{ref12}. On six of the seven
measures, PMSprune did better than DME. DME did better than PMSprune
on specificity. We have subsequently compared PMSprune and DME on
additional datasets such as ChIP-Chip, ChIP-seq, and ABS.

\section{Methods} \label{sec:methods}

Tompa, \emph{et.al.} \cite{ref12} posted a challenge for Motif
Search Problem where each participant was required to predict a
single (or none) motif per dataset even though the 18 data sets have
binding sites for multiple motifs. 13 different algorithms for Motif
Search participated in this challenge.

The results were evaluated using the benchmark data sets available
at \url{http://bio.cs.washington.edu/assessment/} . They have a
collection of 56 datasets. In \cite{ref12} they have tested 13
different programs using several statistical measures. These
programs are AlignACE, ANN-Spec, Consensus, GLAM, The Improbizer,
MEME, MITRA, MotifSampler, Oligo/dyad-analysis, QuickScore,
SeSiMCMC, Weeder and YMF.

We need to define some terms, before the seven statistical measures
for testing the programs can be explained.

\begin{itemize}
  \item \textbf{True Positives}(TP): Number of positions in both known sites as well as predicted
  sites.
  \item \textbf{False Positives}(FP): Number of positions in predicted sites which are not present in known sites
  \item \textbf{True Negatives}(TN): Number of positions which are neither in known sites nor in
predicted sites.
  \item \textbf{False Negatives}(FN): Number of positions that are in known sites but not in predicted
  sites.
  \item \textbf{Sensitivity}(Sn): It represents the fraction of sites that were correctly
  predicted.
  \item \textbf{Sensitivity}(Sp): It represents the fraction of non-sites that were correct
  \item \textbf{Positive Predictive Value}(PPV): It represents the fraction of predicted sites that are
  known.
\end{itemize}

The Sn, Sp and PPV measures are computed as follows (x can either be
n for nucleotide or s for site):

\begin{description}
  \item[1] $xSn = \frac{xTP}{xTP + xFN}$
  \item[2] $xSp = \frac{xTN}{xTP + xFP}$
  \item[3] $xPPV= \frac{xTP}{xTP + xFP}$
\end{description}

To compute the value for $sSn$, we label a site to be a true
positive if there is at least a 25\% overlap between the predicted
and the known sites.

Nucleotide level \textbf{Performance Coefficient} is defined as:
\begin{description}
  \item[4] $nPC = \frac{nTP}{nTP+nFN+nFP}$
\end{description}

Nucleotide level \textbf{Correlation Coefficient} is defined as:
\begin{description}
  \item[5] $NCC = \frac{nTP.nTN - nFN.nFP}{(nTP+nFN)(nTN+nFP)(nTP+nFP)(nTN+nFN)}$
\end{description}

In this paper we compare the performance of PMSprune with the above
13 algorithms (indirectly) and the DME algorithm on the Benchmark
datasets in \cite{ref12}. Since a comparison of the above 13
algorithms are already available, a comparison of PMSprune with
these 13 algorithms has been done using the results reported in
\cite{ref12}

\begin{definition} $ d(x, S_i) = \min_{\forall y \in S_i}d(x, y)$,
where $y$ is an $\ell$-mer in $S_i$ and $d(x,y)$ is Hamming Distance
between $x$ and $y$.
\end{definition}

\begin{description}
  \item[Step 1] Find PMS Motifs. \\
  Let $M$ be the list of predicted Motifs from PMS algorithm such that length of $M$, $|M| = m$.
  \begin{itemize}
    \item PMSSumMin: For each motif $x$ in $M$, we compute the Score as
    follows. $Score = \sum_{1\leq i \leq n}d(x, S_i)$. \\After the score is
    computed for each $x$, the motifs are sorted based on Score. We choose
    a subset $M'$ from PMS motifs using the following techniques. \\Find MinScore = Minimum
    value of score out of Scores of all motifs $x$. \\All motifs that have Score = MinScore are
    added to the list $M'$.
    \item PMSSumMin10: The top 10 motifs based on minimum Score is added to the list
    $M'$.
    \item PMSSumMin500: The top 500 motifs based on minimum Score is added to the list
    $M'$.
    \item PMSSumMinD: For each motif $x$ in $M$, we compute the Score as
    follows. If $(d(x, S_i) \leq d)$, then $ScoreD = ScoreD + d(x,
    S_i)$. \\
    After the score is computed for each $x$, the motifs are sorted based on Score.
    We choose a subset $M'$ from PMS motifs using the following
    techniques. \\ Find MinScoreD = Minimum value of score out of ScoreD of all motifs $x$.
    All motifs that have ScoreD = MinScoreD are added to the list
    $M'$.
    \item PMSSumMinD10: The top 10 motifs based on minimum ScoreD are added to the list
    $M'$.
    \item PMSSumMinD500: The top 500 motifs based on minimum ScoreD are added to the list
    $M'$.
  \end{itemize}
  \item[Step 2] Find DME Motifs \\
  The output from DME are the motifs, corresponding PWMs and a Score for each PWM.
  Since, PMS outputs unique string motifs and DME motifs have some degenerate positions
  in them, we open the DME motifs to generate a list of unique motifs and then
  choose a subset $D'$ from all the DME motifs as follows:
    \begin{itemize}
    \item DMESumMin: Let $x_5, x_6 \dots x_15$ be the number of motifs of length 5 to 15
    respectively in PMSSumMin. We generate the list of DME motifs where the list
    has $x_\ell$ motifs of length $\ell$, where $5\leq \ell \leq 15$.
    \item DMESumMinD: Let $x_5, x_6 \dots x_15$ be the number of motifs of length 5 to 15
    respectively in PMSSumMinD. We generate the list of DME motifs where the list
    has $x_\ell$ motifs of length $\ell$, where $5\leq \ell \leq 15$.
    \item DMESumMin10: The top 10 motifs out of all the motifs from DME are put into
    $D'$.
    \item DMESumMin500: The top 500 motifs out of all the motifs from DME are put into
    $D'$.
    \item BestDMESumMin: The best m motifs out of all the motifs from DME are included in $D'$,
    where $m$ is the number of motifs in PMSSumMin.
    \item BestDMESumMinD: The best m motifs out of all the motifs from DME are included in $D'$,
    where $m$ is the number of motifs in PMSSumMinD.
    \end{itemize}
  \item[Step 3] Compute Statistical Measures. \\
  We compute the 7 statistical measures for each motif in all the 9 files
  generated. For each dataset the computation is done as follows.
  Plant all the motifs occurring in the dataset with a distance specified
  in Table 1 corresponding to the length of the motif.
  Pick one predicted motif at a time
  from the file and plant it in the dataset using the distance
  specified in Table 1 corresponding to the length of predicted
  motif. Find the values of nTP, nFP, nFN, nFP, sTP, sTN, sFP, sFN.
 Use these values to compute the 7 Statistical measures nSn, nSp,
 sSn, nPPV, sPPV, nPC, and nCC.
\end{description}

\begin{table}[h]\label{table:01}
\begin{center}
\begin{tabular}{|c|c|}
  \hline
  \textbf{Length(s)} & \textbf{Hamming Distance} \\ \hline
  5,6 & 1 \\ \hline
  6, 8 & 2 \\ \hline
  9, 10, 11 & 3 \\ \hline
  12, 13, 14 & 4 \\ \hline
  15 & 5 \\ \hline
\end{tabular}
\end{center} \caption{The $(l, d)$ pairs used for testing}
\end{table}

\section{Results and Discussions}\label{sec:result}
\subsection{Result for Tompa's dataset}
We tested PMSprune and DME on ChIP-Chip, ChIP-Seq, ABS and Tompa's
datasets \cite{ref12}. Table 2 shows the results for 13 algorithms
that were initially evaluated  in \cite{ref12}. The data for the
table was obtained from
\url{http://bio.cs.washington.edu/assessment/assessment_result.html}
. The table shows the average values over all the datasets for all
four species. It is observed that ANN-Spec shows the best result
based on Sensitivity which is 0.087 followed by Weeder with
Sensitivity as 0.086. Both PMSprune and DME perform better than
these 13 algorithms on sensitivity and other measures. However, when
it comes to specificity, many of the 13 algorithms perform better.

\begin{table}[h]\label{table:02}
\begin{center}
\begin{tabular}{|l|c|c|c|c|c|c|c|}
  \hline
  \textbf{Algorithm} & \textbf{nSn} & \textbf{nSp} & \textbf{sSn} & \textbf{nPPV} & \textbf{sPPV} & \textbf{nPC} & \textbf{nCC} \\ \hline
  AlignACE & 0.0551 & 0.9914 & 0.0881 & 0.1118 & 0.1226 & 0.0383 & 0.0650 \\ \hline
  ANN-Spec & 0.0870 & 0.9822 & 0.1551 & 0.0881 & 0.0848 & 0.0458 & nan \\ \hline
  Consensus& 0.0205 & 0.9968 & 0.0402 & 0.1132 & 0.1329 & 0.0176 & 0.0404\\ \hline
  GLAM    & 0.0257 & 0.9872  & 0.0459 & 0.0381 & 0.0482 & 0.0156 & nan\\ \hline
  The Improbizer & 0.0685 & 0.9819 & 0.1226 & 0.0695 & 0.0836 & 0.0357 & nan\\ \hline
  MEME   & 0.0670 & 0.9890 & 0.1111 & 0.1071 & 0.1394 & 0.0430 & 0.0706\\ \hline
  MEME3  & 0.0776 & 0.9847 & 0.1245 & 0.0909 & 0.1348 & 0.0437 & nan\\ \hline
  MITRA  & 0.0313 & 0.9906 & 0.0498 & 0.0623 & 0.0623 & 0.0213 & 0.0309\\ \hline
  MotifSampler & 0.0600 & 0.9901 & 0.0977 & 0.1069 & 0.1013 & 0.0399 & 0.0666\\ \hline
  Oligo/dyad-analysis & 0.0398 & 0.9957 & 0.0727 & 0.1542 & 0.1214 & 0.0326 & 0.0694\\ \hline
  QuickScore & 0.0174 & 0.9889 & 0.0325 & 0.0301 & 0.0185 & 0.0111 & 0.0083\\ \hline
  SeSiMCMC  &  0.0611 & 0.9685 & 0.0804 & 0.0369 & 0.0751 & 0.0235 & 0.0493\\ \hline
  Weeder & 0.0863 & 0.9960 & 0.1609 & 0.2996 & 0.2886 & 0.0718 & 0.1523\\ \hline
  YMF & 0.0639 & 0.9920 & 0.1206 & 0.1369 & 0.1200 & 0.0455 & 0.0815\\ \hline
\end{tabular}
\end{center} \caption{Results for 13 algorithms evaluated in Tompa \emph{et. al.} [12]}
\end{table}

\subsection{Results for Tompa's Database for PMSprune and DME}

There are 56 datasets in the database used in \cite{ref12}. Due to
limitations of DME and PMS we could not test all the datasets. PMS
does not work for datasets which have only one sequence in them.
Tables 3 and Table 4 show the 7 Statistical Measures for all the 9
Scoring methods we used. The tables represent the values for the
Best Motif that we found in Mus09r and Hm26r datasets, respectively.
From Table 2 it is observed that although the sensitivity of DME
algorithm is very good PMS performs better than DME in all the
cases. The specificity for DME is better than PMS in most of the
cases but the difference is 1.5\% or less.

\begin{table}[h]\label{table:03}
\begin{center}
\begin{tabular}{|l|c|c|c|c|c|c|c|}
  \hline
  \textbf{Scoring Technique} & \textbf{nSn} & \textbf{nSp} & \textbf{sSn} & \textbf{nPPV} & \textbf{sPPV} & \textbf{nPC} & \textbf{nCC} \\ \hline
  PMS SumMin & 0.6139 & 0.9227 & 0.6149 & 0.3240 & 0.9988 & 0.2692 & 0.4009 \\ \hline
  DME SumMin & 0.4385 & 0.9354 & 0.5380 & 0.2907 & 0.9986 & 0.2118 & 0.3093 \\ \hline
  BestDME SumMin & 0.4385 & 0.9354 & 0.5380 & 0.2907 & 0.9986 & 0.2118 & 0.3093 \\ \hline
  PMS SumMinD & 0.7718 & 0.8910 & 0.7769 & 0.2993 & 0.9986 & 0.2750 & 0.4340\\ \hline
  DME SumMinD & 0.5788 & 0.8899 & 0.6659 & 0.2409 & 0.9983 & 0.2050 & 0.3161\\ \hline
  BestDMESumMinD & 0.4385 & 0.9354 & 0.5380 & 0.2907 & 0.9986 & 0.2118 & 0.3093\\ \hline
  PMS SumMin500 & 0.4386 & 0.9408 & 0.4167 & 0.3125 & 1.0000 & 0.2232 & 0.3251\\ \hline
  PMS SumMinD500 & 0.4386 & 0.9408 & 0.4167 & 0.3125 & 1.0000 & 0.2232 & 0.3251\\ \hline
  DME SumMin500 & 0.0877 & 0.9440 & 0.0833 & 0.0877 & 1.0000 & 0.0459 & 0.0327\\ \hline
\end{tabular}
\end{center} \caption{Result for Dataset Mus09r}
\end{table}

\begin{table}[h]\label{table:04}
\begin{center}
\begin{tabular}{|l|c|c|c|c|c|c|c|}
  \hline
  \textbf{Scoring Technique} & \textbf{nSn} & \textbf{nSp} & \textbf{sSn} & \textbf{nPPV} & \textbf{sPPV} & \textbf{nPC} & \textbf{nCC} \\ \hline
  PMS SumMin & 0.3889 & 0.8745 & 0.4970 & 0.2516 & 0.9999 & 0.1803 & 0.2184 \\ \hline
  DME SumMin & 0.0658 & 0.9192 & 0.1265 & 0.0811 & 0.9995 & 0.0377 & -0.0166 \\ \hline
  BestDME SumMin & 0.1020 & 0.9646 & 0.1383 & 0.2381 & 0.9992 & 0.0769 & 0.0987 \\ \hline
  PMS SumMinD & 0.3889 & 0.8745 & 0.4970 & 0.2516 & 0.9999 & 0.1803 & 0.2184 \\ \hline
  DME SumMinD & 0.0658 & 0.9192 & 0.1265 & 0.0811 & 0.9995 & 0.0377 & -0.0166 \\ \hline
  BestDMESumMinD & 0.1020 & 0.9646 & 0.1383 & 0.2381 & 0.9992 & 0.0769 & 0.0987 \\ \hline
  PMS SumMin500 & 0.3447 & 0.8695 & 0.4678 & 0.2229 & 1.0000 & 0.1565 & 0.1776 \\ \hline
  PMS SumMinD500 & 0.3447 & 0.8695 & 0.4678 & 0.2229 & 1.0000 & 0.1565 & 0.1776 \\ \hline
  DME SumMin500 & 0.0669 & 0.9663 & 0.0851 & 0.1777 & 1.0000 & 0.0511 & 0.0525 \\ \hline
\end{tabular}
\end{center} \caption{Result for Dataset Hm26r}
\end{table}

Table 5 compares the performance of results from all the 9 scoring
techniques. We do the comparison in the following manner: For each
dataset we compare the 7 Statistical Measures for all the 9 scoring
techniques and find out which technique gave the best result for
that measure. Then we count the total number of times the scoring
technique gave the best result for that Statistical Measure. The
numbers in Table 4 represent the count of best performance for each
technique for each measure. The sum of each column is 45 which is
the total number of datasets that we tested.

\begin{table}[h]\label{table:05}
\begin{center}
\begin{tabular}{|l|c|c|c|c|c|c|}
  \hline
  \textbf{Scoring Technique} & \textbf{nSn} & \textbf{nSp} & \textbf{sSn} & \textbf{nPPV} & \textbf{nPC} & \textbf{nCC} \\ \hline
  PMS SumMin & 18 & 0 & 16 & 8 & 13 & 7\\ \hline
  DME SumMin & 0 & 3 & 0 & 3 & 2 & 2\\ \hline
  BestDME SumMin & 1 & 5 & 1 & 1 & 0 & 0\\ \hline
  PMS SumMinD & 4 & 0 & 3 & 4 & 2 & 2\\ \hline
  DME SumMinD & 0 & 7 & 0 & 0 & 1 & 1\\ \hline
  BestDMESumMinD & 0 & 8 & 0 & 2 & 1 & 2\\ \hline
  PMS SumMin500 & 18 & 1 & 20 & 13 & 20 & 11\\ \hline
  PMS SumMinD500 & 4 & 0 & 5 & 6 & 3 & 7\\ \hline
  DME SumMin500 & 0 & 21 & 0 & 8 & 3 & 12\\ \hline
\end{tabular}
\end{center} \caption{Best performers}
\end{table}

Table 6 shows the average values of each statistical measure over
all the datasets for all the 9 scoring techniques. It is observed
that the average value of nucleotide as well as site Sensitivity for
PMS is better than DME. There is at least a difference of 8\% or
more for nSn and 10\% or more for sSn. DME is better with respect to
Specificity. The difference is no more than 3.5\%. PMS is better
than DME based on the other 4 measures as well.

\begin{table}[h]\label{table:06}
\begin{center}
\begin{tabular}{|l|c|c|c|c|c|c|c|}
  \hline
  \textbf{Scoring Technique} & \textbf{nSn} & \textbf{nSp} & \textbf{sSn} & \textbf{nPPV} & \textbf{sPPV} & \textbf{nPC} & \textbf{nCC} \\ \hline
  PMS SumMin & 0.2586 & 0.9164 & 0.3483 & 0.3752 & 0.9997 & 0.1561 & 0.1843 \\ \hline
  DME SumMin & 0.1638 & 0.9464 & 0.2332 & 0.3623 & 0.9994 & 0.1122 & 0.1385 \\ \hline
  BestDME SumMin & 0.1715 & 0.9377 & 0.2419 & 0.3366 & 0.9993 & 0.1122 & 0.1280\\ \hline
  PMS SumMinD & 0.2880 & 0.9163 & 0.3695 & 0.3951 & 0.9996 & 0.1727 & 0.2100\\ \hline
  DME SumMinD & 0.1936 & 0.9420 & 0.2696 & 0.3715 & 0.9995 & 0.1245 & 0.1566\\ \hline
  BestDMESumMinD & 0.1863 & 0.9346 & 0.2626 & 0.3495 & 0.9993 & 0.1220 & 0.1409\\ \hline
  PMS SumMin500 & 0.2322 & 0.8961 & 0.3117 & 0.3809 & 0.9778 & 0.1476 & 0.1674\\ \hline
  PMS SumMinD500 & 0.2349 & 0.8959 & 0.3160 & 0.3842 & 1.0000 & 0.1487 & 0.1687\\ \hline
  DME SumMin500 & 0.1048 & 0.9302 & 0.1485 & 0.2802 & 0.8224 & 0.0747 & 0.0748\\ \hline
\end{tabular}
\end{center} \caption{Average of all datasets for Best Results}
\end{table}

\subsection{ChIP-Chip Results}
We used the publicly available database for ChIP-Chip available at
\url{http://bdtnp.lbl.gov/Fly-Net/browseChipper.jsp} . Since some
datasets are very large we only use the datasets which have less
than 500 sequences and length of motifs are less than or equal to
15. A summary of the results can be found in Table 7.

\begin{table}[h]\label{table:07}
\begin{center}
\begin{tabular}{|l|c|c|c|c|c|c|c|}
  \hline
  \textbf{Scoring Technique} & \textbf{nSn} & \textbf{nSp} & \textbf{sSn} & \textbf{nPPV} & \textbf{sPPV} & \textbf{nPC} & \textbf{nCC} \\ \hline
  PMSSumMin &  0.1464 & 0.8304 & 0.2254 & 0.3799 & 1.0000 & 0.1168 & 0.1463 \\ \hline
  PMSSumMinD & 0.1575 & 0.8148 & 0.2322 & 0.3603 & 1.0000 & 0.1058 & 0.1143\\ \hline
  PMSSumMin500 & 0.1441 & 0.8351 & 0.2253 & 0.3883 & 1.0000 & 0.1169 & 0.0762\\ \hline
  PMSSumMinD500 & 0.1181 & 0.8944 & 0.1848 & 0.4314 & 1.0000 & 0.1003 & 0.0277 \\ \hline
  PMSSumMin10 & 0.1667 & 0.8342 & 0.2481 & 0.4181 & 1.0000 & 0.1364 & 0.112 \\ \hline
  PMSSumMinD10 & 0.1278 & 0.8928 & 0.1890 & 0.4328 & 1.0000 & 0.1089 & 0.1275 \\ \hline
  DMESumMin & 0.1416 & 0.8717 & 0.2025 & 0.4118 & 1.0000 & 0.1193 & 0.0422 \\ \hline
  DMESumMinD & 0.1485 & 0.8741 & 0.2079 & 0.4300 & 1.0000 & 0.1265 & 0.0608 \\ \hline
  DMESumMin500 & 0.1282 & 0.8833 & 0.1918 & 0.4232 & 0.9997 & 0.1077 & -0.02588 \\ \hline
  DMESumMin10 & 0.1546 & 0.8855 & 0.2138 & 0.4553 & 1.0000 & 0.1339 & 0.0475\\ \hline
\end{tabular}
\end{center} \caption{Average Results for ChIP-Chip database}
\end{table}

\subsection{ChIP-Seq Results}
We used the Human database for the ChIP-Seq results. It consisted of
24 files (Chr1 - Chr22, ChrX, ChrY). Since the datasets were large,
we just sampled a small subset of sequences. 20\% of the total
number of sequences were chosen at random and all motifs which occur
in at least half the sequences are chosen by PMS. The motif and the
locations in the sequences are available and we used that
information to find where the truth occurred. The comparison results
are summarized in Table 8.

\begin{table}[h]\label{table:08}
\begin{center}
\begin{tabular}{|l|c|c|c|c|c|c|c|}
  \hline
  \textbf{Scoring Technique} & \textbf{nSn} & \textbf{nSp} & \textbf{sSn} & \textbf{nPPV} & \textbf{sPPV} & \textbf{nPC} & \textbf{nCC} \\ \hline
  PMSSumMin & 0.1554 & 0.8613 & 0.2458 & 0.2086 & 0.9999 & 0.0977 & 0.1554\\ \hline
  PMSSumMinD & 0.1339 & 0.8923 & 0.2142 & 0.2195 & .9999 &  0.08176 & 0.1793\\ \hline
  PMSSumMinD500 & 0.1135 & 0.9061 & 0.1555 & 0.1896 & 0.9998 & 0.0639 & 0.0935 \\ \hline
  PMSSumMin500 & 0.1284 & 0.8802 & 0.2068 & 0.1879 & 0.9999 & 0.0776 & 0.1184 \\ \hline
  DMESumMin & 0.1128 & 0.9121 & 0.1687 & 0.1365 & 0.9989 & 0.0453 & 0.1769 \\ \hline
  DMESumMinD & 0.1043 & 0.8796 & 0.0618 & 0.1170 & 0.9999 & 0.0326 & 0.0429 \\ \hline
  DMESumMin500 & 0.1189 & 0.9361 & 0.1963 & 0.1376 & 0.9998 & 0.0960 & 0.0891 \\ \hline
\end{tabular}
\end{center} \caption{Average Results for ChIP-Seq database}
\end{table}

\subsection{ABS Results}
ABS gives us an option to choose the number of sequences, length of
sequences, the Motif to be planted along with the probability to
plant the motif as an input. We generated datasets of 50 sequences
with length 500 each and planted 1 motif with probability 1. Since,
ABS plants the motif exactly in about 50\% of the sequences, we
generate the list M of PMS motifs by finding all motifs which occur
exactly in 20 out of 50 sequences. Table 9 presents a summary of the
comparison.

\begin{table}[h]\label{table:09}
\begin{center}
\begin{tabular}{|l|c|c|c|c|c|c|c|}
  \hline
  \textbf{Scoring Technique} & \textbf{nSn} & \textbf{nSp} & \textbf{sSn} & \textbf{nPPV} & \textbf{sPPV} & \textbf{nPC} & \textbf{nCC} \\ \hline
  PMS & 0.1934 & 0.9104 & 0.2298 & 0.5275 & 1.0000 & 0.1783 & 0.2168 \\ \hline
  DME & 0.1485 & 0.9148 & 0.1650 & 0.3187 & 1.0000 & 0.1256 & 0.1797 \\ \hline
\end{tabular}
\end{center} \caption{Average Results for ABS database}
\end{table}

\section{Conclusions}
In this paper we have compared PMSprune, one of the best known exact
algorithms based on the $(l, d)$ motif model, with 14 other
algorithms. In particular, we have employed the 13 algorithms
reported in \cite{ref12} and DME. These algorithms have been
compared based on seven different statistical measures. This
comparison shows that PMSprune is very competitive with the other 14
algorithms.

\section{Acknowledgements} \label{sec:ack}
This work has been supported in part by the following grants: NSF
0326155, NSF 0829916 and NIH 1R01GM079689-01A1.

\bibliographystyle{alpha}
\bibliography{paper}

\newcommand{\etalchar}[1]{$^{#1}$}
\begin{thebibliography}{vHRCV00}

\bibitem[ASZ05]{ref13}
A.D.~Smith AD, P.~Sumazin, and M.Q. Zhang.
\newblock Identifying tissue-selective transcription factor binding sites in
  vertebrate promoters.
\newblock In {\em Proceeding of National Academic Science, USA}, 2005.

\bibitem[BT01]{ref05}
J.~Buhler and M.~Tompa.
\newblock Finding motifs using random projections.
\newblock In {\em Proceedings of Fifth Annual International Conference on
  Computational Molecular Biology (RECOMB)}, April 2001.

\bibitem[CL05]{ref11}
F.Y.L. Chin and H.C.M. Leung.
\newblock Voting algorithms for discovering long motifs.
\newblock In {\em Proceedings of the Third Asia-Pacific Bioinformatics
  Conference (APBC2005), Singapore}, pages 261--271, January 2005.

\bibitem[DBR06]{ref09}
J.~Davila, S.~Balla, and S.~Rajasekaran.
\newblock Space and time efficient algorithms for planted motif search.
\newblock In {\em Proceedings of the Second International Workshop on
  Bioinformatics Research and Applications (IWBRA 2006)}, May 2006.

\bibitem[DBR07]{ref03}
J.~Davila, S.~Balla, and S.~Rajasekaran.
\newblock Fast and practical algorithms for planted $(l, d)$ motif search.
\newblock In {\em IEEE/ACM Transactions on Computational Biology and
  Bioinformatics}, pages 544--552, October 2007.

\bibitem[EP02]{ref06}
E.~Eskin and P.~Pevzner.
\newblock Finding composite regulatory patterns in dna sequences.
\newblock {\em Bioinformatics}, S1:354--363, 2002.

\bibitem[ES03]{ref21}
P.~Evans and A.~Smith.
\newblock Toward optimal motif enumeration.
\newblock In {\em Proceeding of WADS}, 2003.

\bibitem[FHSW04]{ref16}
M.C. Frith, U.~Hansen, J.L. Spouge, and Z.~Weng.
\newblock Finding functional sequence elements by multiple local alignment.
\newblock {\em Nucleic Acids Research}, 2004.

\bibitem[KLK{\etalchar{+}}01]{ref17}
S.K. Kim, J.~Lund, M.~Kiraly, K.~Duke, M.~Jiang, J.M. Stuart, A.~Eizinger, B.N.
  Wylie, and G.S. Davidson.
\newblock A gene expression map for caenorhabditis elegans.
\newblock {\em Science}, 29(3), 2001.

\bibitem[PMMP04]{ref22}
G.~Pavesi, P.~Mereghetti, G.~Mauri, and G.~Pesole.
\newblock Weeder web: discovery of transcription factor binding sites in a set
  of sequences from co-regulated genes.
\newblock {\em Nucleic Acids Research}, 32, 2004.

\bibitem[PRP03]{ref08}
A.~Price, S.~Ramabhadran, and P.~A. Pevzner.
\newblock Finding subtle motifs by branching from sample strings.
\newblock {\em Bioinformatics}, 1(1):1--7, 2003.

\bibitem[PS00]{ref07}
P.~Pevzner and S.-H. Sze.
\newblock Combinatorial approaches to finding subtle signals in dna sequences.
\newblock In {\em Proceedings of Eighth International Conference on Intelligent
  Systems for Molecular Biology}, pages 269--278, 2000.

\bibitem[RBH05]{ref01}
S.~Rajasekaran, S.~Balla, and C.-H. Huang.
\newblock Exact algorithms for planted motif challenge problems.
\newblock {\em Journal of Computational Biology}, 12(8):1117--1128, 2005.

\bibitem[RD04]{ref20}
M.~Régnier and A.~Denise.
\newblock Rare events and conditional events on random strings.
\newblock {\em Discrete Mathematics Theory Computer Science}, 6, 2004.

\bibitem[RHEC98]{ref14}
F.P. Roth, J.D. Hughes, P.W. Estep, and G.~Church.
\newblock Finding dna regulatory motifs within unaligned noncoding sequences
  clustered by whole-genome mrna quantitation.
\newblock {\em Nature Biotechnology}, 1998.

\bibitem[ST03]{ref23}
S.~Sinha and M.~Tompa.
\newblock Ymf: a program for discovery of novel transcription factor binding
  sites by statistical overrepresentation.
\newblock {\em Nucleic Acids Research}, 31, 2003.

\bibitem[TLB{\etalchar{+}}]{ref12}
M.~Tompa, N.~Li, T.L. Bailey, G.M. Church, B.~De Moor, E.~Eskin, A.V. Favorov,
  M.C. Frith, Y.~Fu, W.J. Kent, V.J. Makeev, A.A Mironov, W.S. Noble,
  G.~Pavesi, G.~Pesole, M.~Regnier, N.~Simonis, S.~Sinha, G.~Thijs, J.~van
  Helden, M.~Vandenbogaert, Z.~Weng, C.~Workman, C.~Ye, , and Z.~Zhu.
\newblock Assessing computational tools for the discovery of transcription
  factor binding sites.
\newblock {\em Nature Biotechnology}, 23(1).

\bibitem[vHRCV00]{ref19}
J.~van Helden, A.F. Rios, and J.~Collado-Vides.
\newblock Discovering regulatory elements in noncoding sequences by analysis of
  spaced dyads.
\newblock {\em Nucleic Acids Research}, 28, 2000.

\bibitem[WS00]{ref15}
C.T. Workman and G.D. Stormo.
\newblock Ann-spec: A method for discovering transcription factor binding sites
  with improved specificity.
\newblock In {\em Proceeding of Pacific Symposium on Biocomputing}, 2000.

\end{thebibliography}

\end{document}